# Heat Conductance of the Quantum Hall Bulk

Ron Aharon Melcer[1,2], Avigail Gil[2], Arup-Kumar Paul[1,2], Priya Tiwary[1,2],

Vladimir Umansky[1,2], Moty Heiblum[1,2], Yuval Oreg[2], Ady Stern[2], and Erez Berg[2]

[1] Braun Center for Submicron Research, Department of Condensed Matter Physics,

 Weizmann Institute of Science, Rehovot, Israel, 76100

[2] Department of Condensed Matter Physics, Weizmann Institute of Science, Rehovot, Israel, 76100

**The Quantum Hall Effect (QHE) is a prototypical realization of a topological state of matter. It emerges from a subtle interplay between topology, interactions, and disorder. The disorder enables the formation of localized states in the bulk that stabilize the quantum Hall states with respect to the magnetic field and carrier density. Still, the details of the localized states and their contribution to transport remain beyond the reach of most experimental techniques. Here, we describe an extensive study of the bulk's heat conductance. Using a novel 'multi-terminal' short device (on a scale of $10\mu m$), we separate the longitudinal thermal conductance, $\kappa_{xx}T$ (due to bulk's contribution), from the topological transverse value $\kappa_{xy}T$, by eliminating the contribution of the edge modes. When the magnetic field is tuned away from the conductance plateau center, the localized states in the bulk conduct heat efficiently ($\kappa_{xx}T \propto T$), while the bulk remains electrically insulating. Fractional states in the first excited Landau level, such as the $\nu = 7/3$ and $\nu = 5/2$, conduct heat throughout the plateau with a finite $\kappa_{xx}T$. We propose a theoretical model that identifies the localized states as the cause of the finite heat conductance, agreeing qualitatively with our experimental findings.**

The magneto-conductance trace is the best-known signature of the quantum Hall effect (QHE). At integer (IQHE) or simple fractional (FQHE) values of the Landau-Levels (LL) fillings $\nu$, the electrical conductance differs from the classical predictions. The transverse conductance exhibits plateaux (as a function of magnetic field or carrier density) with quantized values, $G_{xy} = \nu G_0$ ($G_0 = \frac{e^2}{h}$), while the longitudinal conductance vanishes, $G_{xx} \sim 0$. This observation implies that the bulk is insulating and current can only flow next to the sample's edge [1, 2]. Due to disorder, the localized states in the bulk get filled or emptied once the magnetic field is tuned away from the plateau's center. These localized states are a crucial ingredient for stabilizing the QHE states and a classical Hall conductance (i.e., without quantized plateaux in $G_{xy}$) is expected to re-emerge in perfectly clean samples [3]. The contribution of the localized states to the electrical transport is negligible, since any deviation of the electrical conductance from its temperature-activated behavior - $G_{xx} \propto \exp\left(-\frac{\Delta}{T}\right)$, with $\Delta$ the energy gap and $T$ the temperature - is hard to analyze quantitatively due to its smallness in the limit of $T \to 0$ [4, 5, 6].

The thermal Hall conductance is a second transport coefficient that is of great importance for studying the QHE. Similar to $G_{xy}$, the thermal Hall conductance is also quantized for QHE states[7, 8], $\kappa_{xy}T = \nu_Q \, \kappa_0 T$, where $\kappa_0 = \frac{\pi^2 k_b^2}{3h}$ is the thermal conductance quantum ($T$ - temperature, $k_b$ - Boltzman constant, $h$ - Planck's constant), and $\nu_Q$ depends on the nature of the topological order. It is quantized to an integer for all Abelian states (integer and most fractional filling fractions) and a fraction for a non-Abelian state[9]. The value of $\nu_Q$ is a distinct property of the topological order of the bulk, which can provide crucial insight into the underlying ground state. The study of thermal transport in the QHE has accelerated in recent years, with experimental results manifesting the quantization of the *two-terminal* thermal conductance coefficient, $\kappa_{2T}$ [10, 11, 12, 13]. These studies shed new light on the nature of fractional QHE states, especially on the exotic $\nu = 5/2$ state [11, 14]. However, $\kappa_{2T}$ is sensitive to the inter-mode thermal equilibration and a possible bulk contribution [15, 16, 17].

Following the fact that $G_{xx}$ vanishes on the QHE plateaux, it is tempting to assume that the bulk is also thermally insulating (despite some indirect evidence of possible heat flow through the bulk[18, 19, 20]). A possible justification for such simplification is due to the Wiedemann-Franz (WF) conjecture, which agreed experimentally in metallic systems with a fixed ratio $\kappa_{xx}/G_{xx} = \kappa_0/G_0 = L_0$, known as the 'Lorenz number'. Substantial violations of $L_0$ appeared in the limit $T \to 0$ in systems where a single electron picture is no longer valid, and the electronic transport is affected by interactions [21, 22, 23].

Recently, we developed a novel technique that enables a local measurement of propagating power [24], thus allowing a direct determination of the transverse thermal Hall conductance, $\kappa_{xy}T$. Here we employed this technique to determine the heat flow through bulk, using the term 'longitudinal thermal conductance' or $\kappa_{xx}T$. We found finite $\kappa_{xx}$ at QHE conductance plateaus (where $G_{xx}$, $R_{xx}$~0), and propose a dominant heat transport mechanism.

The employed configuration, shown in Fig. 1, was fabricated in a 2DEG within a GaAs-AlGaAs heterostructure (see details in Methods). It consists of two floating metallic reservoirs, a 'temperature source' (S), and a 'power meter' (PM) located on the other side of the mesa, some 10μm away (see also Ref.[24] and Methods). The source S is heated by injecting two opposite-sign currents, $I$ and $-I$, emanating from two separate contacts, $S_1$ and $S_2$ outside the active area of the short mesa, leading to zero source potential. The power is evacuated from the source via the outgoing gapless edge modes and possibly the bulk. Here, the edge modes flow to side-grounded contacts, thus not reaching the PM. This configuration isolates the heat flow through the bulk, emanating from the source contact or from the diverted edge modes (dashed arrows in Fig. 1 illustrate this). The temperatures of S and PM are determined by measuring the emanating Johnson-Nyquist (JN) noise reaching the amplifiers' contacts, $A_S$ and $A_{PM}$, respectively. The PM's elevated temperature is converted to the absorbed power using a separate calibration measurement (see Methods).

The longitudinal thermal conductance coefficient is extracted from the dissipated power in the PM and the temperature of the source,

$$P = \gamma \frac{\kappa_{xx}(T)}{2}(T_S^2 - T_0^2), \qquad (1)$$

where $T_S$ and $T_0$ are the source and the base temperatures (respectively), $T$ their mean, and $\gamma$ a geometric pre-factor (in our device being of the order of one). The bulk's electrical conductance, $G_{xx}$, is measured by applying a voltage to the source contact and measuring the current drained through the PM contact (see Methods). In the rest of the manuscript, we simplify the expressions by absorbing the geometric pre-factor into the definition of $\kappa_{xx}$ and $G_{xx}$ (namely taking $\gamma = 1$). We emphasize that the reported values of $\kappa_{xx}$ are not universal (as they depend on the dimensions of the device), while the ratio $\kappa_{xx}/G_{xx}$ is a general property of the bulk and thus independent of its geometry.

We start with filling factor $\nu = 2$. The power arriving at the PM is plotted as a function of the source's temperature at three different locations on the conductance plateau (Fig. 2a): (*i*) Center, $\nu = 2$ ($B = 6.4T$) - the first spinless Landau level is fully occupied; (*ii*) $\nu \sim 2.05$ ($B = 6.1T$) - localized states in the upper spinful LL2 are partly occupied; and (*iii*) $\nu \sim 1.95$ ($B = 6.7T$), localized quasi-holes in the spinful LL1 are partly occupied. We find a vanishingly small bulk heat conductance in the plateau's center, guaranteeing no 'parallel' mechanism that efficiently carries heat from S to PM (such as phonons or mobile carriers in the doping layer [25]). However, a finite heat conductance is observed when the field is tuned away from the plateau's center.

The thermal ($\kappa_{xx}T$) and electrical ($G_{xx}$) conductances are plotted as a function of the magnetic field on $\nu = 2$ plateau (Fig. 2b). We observe a narrow region of the magnetic field (near the center of the plateau) where $\kappa_{xx}$ vanishes. However, it increases rapidly towards the edges of the plateau, thus violating the WF relation by as much as two orders of magnitude, $\kappa_{xx}/G_{xx} \approx 200L_0$ for $B = 5.9T$.

The plotted temperature dependence of the transmitted power for three different fillings (Fig. 3a) suggests temperature-independent $\kappa_{xx}$ in the range of $T_0 = 15\text{--}60\,mK$. In addition, extending the temperature range of the source to $T_S = 20\text{--}100\,mK$ (thus creating large thermal gradients across the device) was also found to be consistent with temperature-independent $\kappa_{xx}$ (Fig. 3b). This implies that the thermal conductance, $\kappa_{xx}T \propto T$, in a similar fashion to a metal at low temperatures [26].

We now discuss a simple physical picture that may explain the origin of the bulk thermal conductance in the QHE and its temperature dependence. In the presence of potential disorder, tuning the band filling away from the center of the conductance plateau adds localized quasi-particles (or quasi-holes). For a relatively smooth disorder potential (relative to the magnetic length), the added quasi-particles form incompressible "puddles" with fillings different from that of the bulk (see Fig. 4a). In the $\nu = 2$ bulk filling, added quasi-holes may form puddles with filling $\nu = 1$, with boundaries supporting gapless chiral edge modes. While particle tunneling between these puddles decays exponentially with the distance between the puddles, Coulomb interaction decays in a power law fashion[27]. We assume that the puddles are large enough so that the level-spacing of excitations within a typical puddle is smaller than the temperature (we elaborate on this condition below).

We make several simplifying assumptions to derive the thermal current's temperature dependence. First, we assume that the thermal conductance between the puddles is small enough to define a local temperature in each puddle. Second, the interaction region between two neighboring puddles is a segment of characteristic length $L_{int}$ (around the point where they are closest), with interaction of a density-density form (illustrated in Fig. 4b). More details of the model, including the explicit form of the Hamiltonian and the derivation of the thermal conductance, are given in the Supplementary Information (SI).

The low-energy excitations along the edge of each puddle are linearly dispersing chiral charge density waves (plasmons) that behave as free bosons. Under the assumptions above, the transmission coefficient $\mathcal{T}(\omega)$, defined as the probability of an incident excitation with frequency $\omega$ transmitted from one puddle to the next [28] (see SI),

$$\mathcal{T}(\omega) = \frac{\frac{u^2}{\tilde{v}^2}\sin^2\frac{\omega L_{int}}{\tilde{v}}}{1 + \frac{u^2}{\tilde{v}^2}\sin^2\frac{\omega L_{int}}{\tilde{v}}}. \qquad (2)$$

Here $u$ is the strength of the inter-edge interaction and $\tilde{v} = \sqrt{v^2 - u^2}$ is the renormalized speed of the excitations in the segment where the edges interact, and $v$ is the speed of excitations along the puddle's edge away from the interaction region (for simplicity, $v$ is taken to be the same in the two puddles). Notice that $u$ is measured in velocity units and must satisfy $u < v$ (see SI). From this expression, the heat current flowing from the 'hot puddle' to the 'colder puddle' is,

$$J_Q(T_L, T_R) = v \int_0^\infty \frac{dk}{2\pi} \hbar \omega_k \left[ n_B \left( \frac{\hbar \omega_k}{k_b T_L} \right) - n_B \left( \frac{\hbar \omega_k}{k_b T_R} \right) \right] \mathcal{T}(\omega_k) , \qquad (3)$$

where $T_L, T_R$ are the temperatures of the two puddles, $\omega_k = vk$ and $n_B(x)$ is the Bose function. In the limit $k_b T \gg \hbar v / L_{int}$, the thermal current simplifies to,

$$J_Q(T_L, T_R) = \tilde{\kappa}(T_L^2 - T_R^2), \qquad (4)$$

where, for $u \ll v$, $\tilde{\kappa} = \frac{\kappa_0}{4} \frac{u^2}{v^2}$. In the opposite limit, $k_b T \ll \hbar v / L_{int}$, we find $J_Q \propto T^4$ [29, 30]. Taking $L_{int} \sim 1\mu m$ and $v \sim 10^4 m/s$ (for a weak electric field at the boundaries), we obtain a characteristic temperature scale $T^* = \frac{\hbar v}{k_b L_{int}} \sim 75mK$, being of the order of the temperature in our experiments. Figure 4c depicts the thermal conductance between two puddles evaluated from Eq. 3. We find that the thermal conductance is approximately temperature-independent already for $T \gtrsim 0.1T^*$ (i.e., according to the estimate above, for $T \gtrsim 7.5mK$).

The model reproduces the temperature dependence of the thermal current observed experimentally. Despite its simplicity, the essential ingredients leading to our main conclusion (Eq. 4) are robust. Heat is transferred through the scattering of plasmon excitations by inter-edge interactions with a transmission coefficient, $\mathcal{T}(\omega)$. When averaged over an energy window of the order of the temperature, it is nearly temperature-independent since the oscillations of $\mathcal{T}(\omega)$ are averaged out at a sufficiently high temperature. For $v$ of the same order of magnitude as $u$, we obtain $\tilde{\kappa} \sim \kappa_0$. The actual value of $\kappa_{xx}$, probed in the experiment, is non-quantized and depends on the device geometry as well as the microscopic realization of the disorder which will fix $\tilde{\kappa}$.

While we cannot rule out other sources for low-energy bulk neutral modes, such as collective modes of various symmetry-broken states (including Wigner crystals and FQH nematic states [31], it is not apparent how such modes can explain the observed and calculated temperature dependence of the thermal current (see SI).

The second Landau level hosts several more fragile fractional states [32, 33, 34], and among them we studied the $\nu = 7/3$ and the $\nu = 5/2$ states. The $\nu = 7/3$ state is believed to be a Laughlin fractional state on top of a full spinless first Landau level [35]. The $\nu = 5/2$ is the only (well-developed) even-denominator state [36]. Its underlying order in the bulk has been the center of intensive experimental and theoretical study in the past two decades[37]. The most decisive identification comes from thermal transport experiments [11, 14], which support a topological order known as Particle-Hole Pfaffian (PH-Pf)[38].

The PH-Pf ground state is inconsistent with current numerical simulations supporting a different non-Abelian order [39, 40].

Our device exhibits Hall resistance plateaus at $\nu = 7/3$ and $\nu = 5/2$ (Fig. 5a), with a vanishingly small bulk conductance at the center of the conductance plateau (as low as $G_{xx} \sim 5 \times 10^{-3} G_0$). On the other hand, the bulk's thermal conductance coefficient, $\kappa_{xx}$, remains finite throughout the plateau, never below $\kappa_{xx} = 0.3\kappa_0$ (two-orders-of-magnitude violation of the WF relation, see Figs. 5b & 5c). We attribute this observation to the small energy gap of these states, leading to excited quasi-particles (and quasi-holes), even at the plateau's center.

In long devices (100 - 200 microns in length), fabricated on similar-quality GaAs, the expected value of the two-terminal thermal conductance was confirmed (e.g., $\kappa_{2T} = 3\kappa_0$ for $\nu = 7/3$) [11-14]. This implies that the quantized heat flows near the sample's edge with a negligible bulk contribution. Indeed, we find that as the devices become longer (smaller $\gamma$ in Eq. 1), the contribution of the bulk heat flow diminishes inversely proportional to its length (see more details SI).

Very little is known about the properties of the localized states in the electrically insulating 2D bulk of topologically non-trivial 2D materials. Studying the heat flow through the bulk of QHE states [20], we find a substantial heat conductance in the integer and fractional QHE states (in particular away from the center of the conductance plateaux) and attribute it to interacting circulating currents around incompressible, disorder-induced, charged puddles. Deviating further from the center of the conductance plateau increases the heat flow, which we associate with the increased number and size of the puddles. Moreover, we find that the longitudinal thermal conductance coefficient $\kappa_{xx}$ (through the bulk) is temperature independent in the measured temperature range 15-100mK. Bulk thermal conductance is an essential ingredient in thermal conductance studies. A deeper understanding is required to determine its effect on topological thermal Hall conductance ($\kappa_{xy}$).

## Methods

### Sample preparation

The sample was fabricated in GaAs-AlGaAs heterostructure. The doping method was double-sided SPSL method [45] with two spaces, each 82nm wide, and a quantum well width of 29nm hosting the 2DEG. The 2DEG density was 2.9x10[11] cm[-2] with dark mobility at 4.2K ~10[7] cm[2]/V-Sec.

The mesa is etch-defined, formed by optical lithography and wet-etching (using the solution $H_2O: H_2O_2: H_2PO_4$ - 50:1:1). The narrow etched lines were separately patterned using e-beam lithography and etched by reactive-ion-etching (RIE) using $BCl_3/Ar$ gases. The Ohmic contacts and the gates were patterned using the standard e-beam-liftoff technique. The ohmic contacts consist of (from GaAs surface upwards) Ni(7nm), Au(250nm), Ge(125nm), Ni (82nm), Au (10nm), alloyed at $440°C$ for 80 seconds. After preparing the contacts, the sample was covered with 25nm of $HfO_2$ deposited at $200°C$ using ALD. The gates consist of 5nm Ti and 15nm Au.

The device has three different gate patterns: (*i*) Thin gates (appear in gray in Fig. 1) that enable local depletion of the carriers. Once activated, they can force the edge modes to flow from S to PM. These gates were grounded throughout the measurements, guaranteeing that the edge modes flow to the ground. (*ii*) Global gates that cover the regions to the left of S and the right of the PM (yellow color in Fig. 1). These gates are used to induce a $\nu = 2$ filling factor in these regions. Note that the measured transport only occurs in the un-gated region between S and PM. The gated regions support Sources' DC currents and Johnson-Nyquist noise. The integer filling offers simplifications to the process of calibration of amplification gain and $T_0$ (see Supplementary information). (*iii*) Grounded gates that cover S and PM contacts (the leads grounding these gates appear in gray in Fig. 1). The 'ohmic contact-oxide-gate' forms a capacitor ($c \sim 0.5 pF$). The enhanced capacitance ensures that the contacts behave as ideal metallic reservoirs[41]. The $HfO_2$ is etched from the bonding pads using RIE ($BCl_3/Ar$ gases).

### Probing $\kappa_{xx}$ and $G_{xx}$

The longitudinal electrical conductivity, $\sigma_{xx}$, and thermal conductivity, $k_{xx}T$, are specific transport properties of the bulk. These properties are usually measured indirectly, using probes placed along the sample's edge (for example, in a Hall-bar geometry). Such methodology allows measuring the global electrical ($G_{xx}$), and thermal ($\kappa_{xx}T$) conductances. For 2D materials, the specific and the global coefficients are equal, up to a device-specific geometric factor known as the number of squares, $N_\square$ . Namely,

$$\sigma_{xx} = G_{xx} \times N_\square$$
$$k_{xx} = \kappa_{xx} \times N_\square.$$

(M1)

Thus the conductance is usually reported for a square. Here we adopted a different approach (described in detail in the main text and the sections below), which is conceptually more suitable for measuring the longitudinal thermal conductance in QHE states. The main problem of measuring the temperature drop along an edge segment (for example, in a Hall-bar geometry) is that it can be sensitive to the thermal conductivity of the bulk (which we want to probe) as well as properties of the edge modes themselves, mostly thermal equilibration (when counter-propagating modes are present)[15, 42, 43], and dissipation of heat from the edge to the environment[43, 44]. We overcame these complications by grounding the edge modes and measuring the heat flow through the bulk directly.

A possible drawback of our technique is that we are unable to determine the number of squares precisely. Thus, the values of $\kappa_{xx}$ reported in the main text are equal to the longitudinal conductivity up to a multiplying pre-factor of order unity (deoted $\gamma$ in Eq. 1). Crucially, we measured the longitudinal electrical conductance using the same contacts as for the thermal conductance (S and PM). Using the same geometry guarantees that the observed WF ratio of the thermal and electrical conductances is purely a bulk property. Moreover, it allows us to rule out the possibility that the observed violations of the WF relation are simply an artifact caused by a 'dirty' region or some density inhomogeneity.

**Electrical conductance measurements**

We use a low-frequency signal (11–17Hz) with a lock-in technique for conductance measurements. We source current $I_S$ from $S_1$ to the source contact, which causes the voltage of the source to be,

$$V = \frac{I_S}{G_{S \to G}}, \tag{M2}$$

with $G_{S \to G}$ the overall conductance from the source to the ground. To measure the longitudinal conductance, we need to measure the current that flows through the bulk, $G_{xx} = I/V$. We probe the source-to-PM current, $I$, via the measurement of the PM voltage (in a contact located in one of the regions to the right of the PM),

$$I = V_{PM} \, G_{PM \to G}, \tag{M3}$$

with $G_{PM \to G}$ the conductance from the PM to the ground. This allows us to extract the longitudinal conductance,

$$G_{xx} = \frac{I}{V} = \frac{V_{PM}}{I_S} G_{S \to G} G_{PM \to G}. \tag{M4}$$

Relying on the fact that $G_{xy} \gg G_{xx}$, we consider only the contribution of the edge modes to the electrical conductance $G_{S \to G}$ and $G_{PM \to G}$. For example, when all the regions are at $\nu = 2$, the conductance from either the PM or the Source to ground is simply the number of 2-DEG regions connected to the contact times the quantized conductance,

$$G_{S \to G} = G_{PM \to G} = 3 \times \frac{2e^2}{h}, \tag{M5}$$

leading to the conductance according to,

$$G_{xx} = 36 \left( \frac{e^2}{h} \right)^2 \frac{V_{PM}}{I_S}. \tag{M6}$$

**Thermal conductance measurements**

Probing the longitudinal thermal conductance requires measuring the power impinging upon the PM as a function of the source's temperature. We accomplish this measurement in two steps: (*i*) We heat the source contact (S) using two opposite polarity DC currents and measure the temperature of the power-meter (PM), $T_{\mathrm{PM}}$, as a function of the temperature of the source $T_{\mathrm{S}}$. (*ii*) We calibrate the PM by measuring its temperature $T_{\mathrm{PM}}$ against a known heating power $P_{\mathrm{cal}}$. We accomplish this by dissipating a known power using opposite sign currents $I_{\mathrm{cal}}$ and $-I_{\mathrm{cal}}$ sourced from contacts $S_1^{\mathrm{cal}}$ and $S_2^{\mathrm{cal}}$ (in Fig. 1) respectively. This causes the direct dissipation of power $P_{\mathrm{cal}} = I_{\mathrm{cal}}^2 / G_{\mathrm{xy}}$ on the PM (here, the source contact remains cold). The measurement of $T_{\mathrm{PM}}$ against $P_{\mathrm{cal}}$ manifests the power-meter calibration.

Combining the primary measurement with the calibration enables us to determine the impinging power on the PM as a function of $T_S$ (see SI for further discussion). We acquire $\kappa_{\mathrm{xx}}$ by linear fitting the power against $T_S^2$, according to Eq. 1.

Once we established that $\kappa_{\mathrm{xx}}$ is temperature independent, we were able to improve the efficiency of our measurement by measuring the power only for a single source temperature $T_{\mathrm{S}}$ (instead of scanning and fitting, which requires significantly longer averaging time). This method was applied for the measurement of $\nu = 5/2$ (Fig. 5c) with $T_{\mathrm{S}} = 40\mathrm{mK}$ (we performed a full scan only in the center of the plateau); for $\nu = 3$, with $T_{\mathrm{S}} = 50\mathrm{mK}$; and for $\nu = 4/3$ (see SI).

# Acknowledgments

R.A.M thanks Alexander D. Mirlin for fruitful discussions. A.G and E.B acknowledge support from the Israel Science Foundation Quantum Science and Technology grant no. 2074/19 and CRC 183 of the Deutsche Forschungsgemeinschaft. M.H acknowledges the support of the European Research Council under the European Union's Horizon 2020 research and innovation program (grant agreement number 833078). A.S acknowledges the support from Israeli Science Foundation Quantum Science and Technology grant no. 2074/19, the CRC 183 of the Deutsche Forschungsgemeinschaft (Project C02), and the European Research Council (ERC) under the European Union's Horizon 2020 research and innovation programme [grant agreements No. 788715 and 817799, Projects LEGOTOP]

# Author contribution

R.A.M designed the experiment, fabricated the devices, performed the measurements and analyzed the data presented in the main text. A.K.P and P.T performed the length dependent measurements presented in SI. M.H supervised the experiment's design, execution, and data analysis. A.G, E.B, Y.O and A.S developed the theoretical model. V.U grew the GaAs heterostructures. All authors contributed to the writing of the manuscript.

# Figures

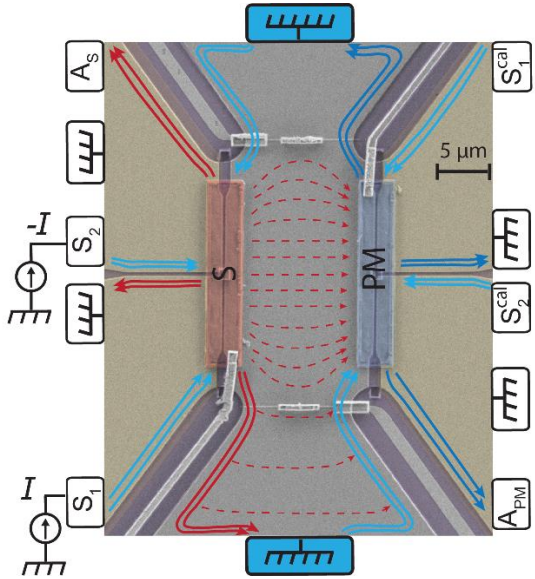

**Figure 1 – The structure of the device.** A false-colored SEM image of the device's heart, which was used to measure the longitudinal thermal conductance, $k_{xx}T$. The mesa is divided into several regions by narrowly etched areas (purple). Heat transport is carried in the central region between the source floating contact (S, colored red) and the power-meter flowating contact (PM, colored dark blue). In the experiment, we heat the source contact by sourcing currents $I$, $-I$ from $S_1$ and $S_2$, respectively. The edge modes (solid lines) that leave S flow to the ground, allowing heat transfer from S to the PM only via the bulk (narrow dashed red lines). The energy currents leaving S reach the PM and increase its temperature. Both temperatures, $T_{PM}$ and $T_S$, are probed simultaneously by measuring the Johnson-Nyquist noise reaching $A_{PM}$ and $A_S$, respectively. The increased temperature of the PM is 'converted' to the impinging power using a calibration measurement (see Methods). The regions to the left of S and the right of the PM (used for noise measurement and control) are fully covered with top gates that tune the areas to the filling $\nu = 2$ (faint yellow), which eases the calibration of the temperature and the power (see SI).

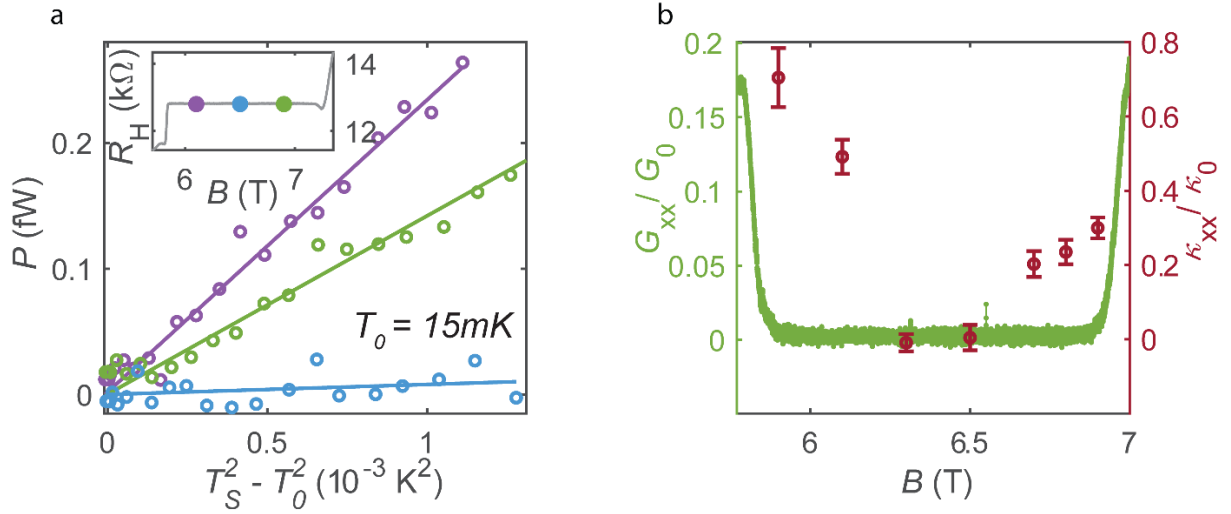

**Figure 2 – Longitudinal thermal conductance $\kappa_{xx}T$ in filling factor $\nu = 2$ at base temperature $T_0 = 15\text{mK}$. (a)** Impinging power at the power meter (PM) as a function of the source (S) temperature at three different values of the magnetic field all on the $\nu = 2$ plateau: Blue – center; purple – particle side; green – hole side (see inset). Away from the center, finite heat flows from S to PM (through the bulk of the sample). By linear fitting the absorbed PM power as a function of its temperature squared (straight colored lines), $\kappa_{xx}$ is extracted (Eq. 1). **(b)** Longitudinal electrical and thermal conductances on the $\nu = 2$ plateau. The longitudinal conductance, $G_{xx}$, vanishes throughout the width of the plateau (green). The thermal conductance coefficient $\kappa_{xx}$, extracted from the linear fits in **(a)**, is finite away from the plateau's center.

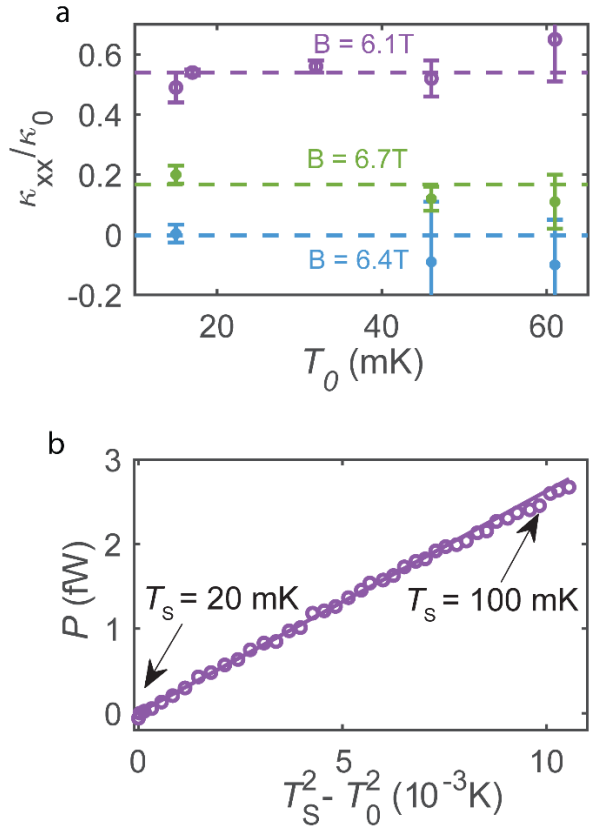

**Figure 3 – Temperature dependence of the longitudinal thermal conductance coefficient $\kappa_{xx}$ at $\nu = 2$.**
**(a)** The longitudinal thermal conductance measured as a function of base temperature, $T_0$, extracted by linear fitting the impinging power at the PM as a function of source temperature squared (see SI). $\kappa_{xx}$ appears to be temperature independent for a few magnetic field values on the plateau (see Fig. 2b). **(b)** Power impinging on the power-meter (PM) as a function of the source (S) temperature for a large temperature range, $T_S = 19 - 100$ mK. The fitted line to the data remains linear up to $T_S = 80$ mK.

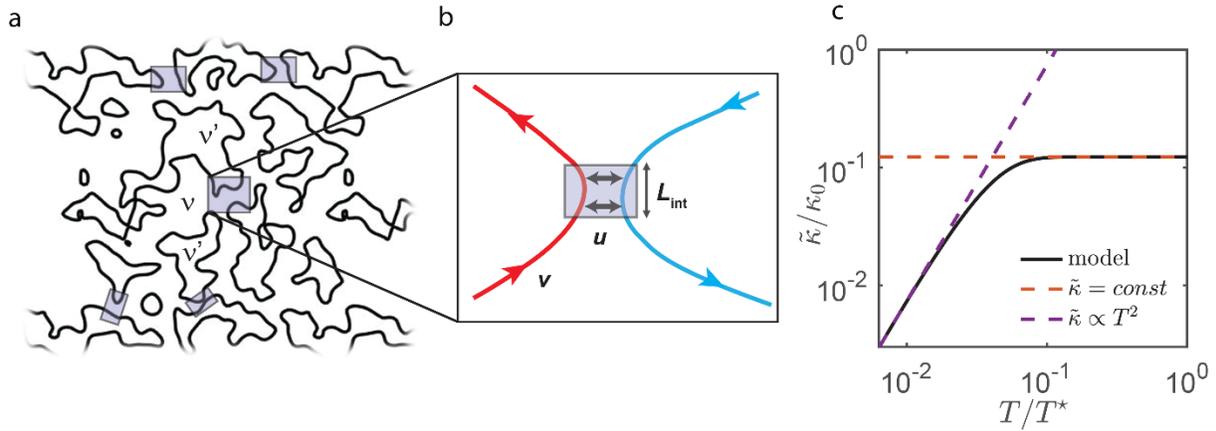

**Figure 4 – Illustration of the model used to calculate the bulk's thermal conductance.** **(a)** Density puddles at filling factor $\nu'$ inside a bulk state at filling factor $\nu$, formed in the presence of smooth disorder when the magnetic field is away from the center of the conductance plateau. Chiral edge modes circulate at the boundaries of the puddles. **(b)** Diagram of puddle-puddle interaction: The distance between two neighboring puddles is large enough to suppress inter-puddle electron tunneling but small enough to allow significant Coulomb interactions. The interactions are of constant strength $u$ and take place in a finite region with a characteristic length of $L_{int}$ (where the edges are the closest). The velocity of the excitations away from the interacting region, $v$, is assumed to be constant and equal in the interacting puddles. **(c)** Thermal conductance coefficient between two interacting puddles. The right puddle is at temperature $T$, and the left is at temperature $T + dt$ (with $dt \ll T$). The thermal conductance coefficient between the puddles is calculated as $\tilde{\kappa} = J_Q/2k_b T dt$, with $J_Q$ according to Eq. 3. For $T \sim 0.1 T^\star$ ($T^\star = \hbar/v k_b L_{in} \approx 75$mK), $\tilde{\kappa}$ saturates and becomes temperature independent, in agreement with the experimental observations, for $u/v = 0.1$.

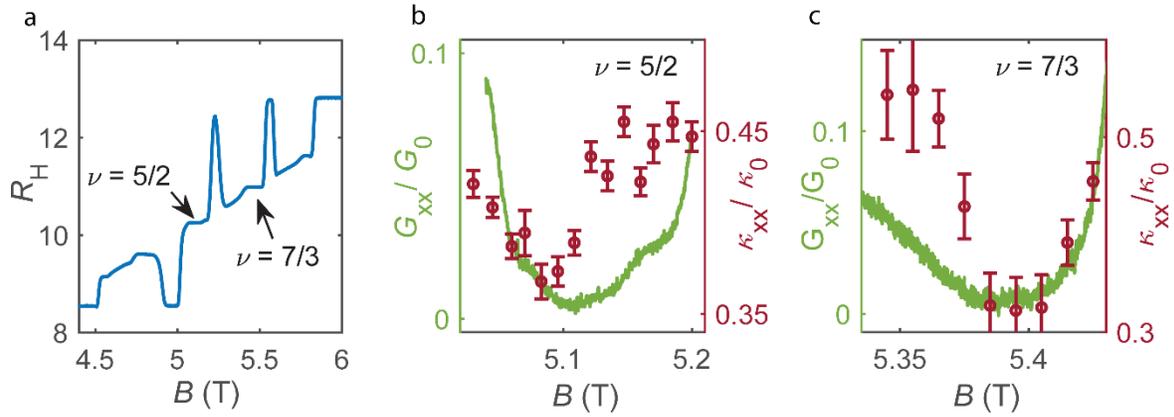

**Figure 5 – Thermal conductance of the bulk at fillings $\nu = 7/3$ and $\nu = 5/2$. (a)** The transverse Hall resistance as a function of the magnetic field manifests plateaus at fractional filling factors in the range between $\nu = 2$ and $\nu = 3$. **(b, c)** $G_{xx}$ (green) and $\kappa_{xx}$ (red) as a function of the magnetic field along the $\nu = 5/2$ **(b)** and $\nu = 7/3$ **(c)** plateaus. Close to the center of each plateau, $G_{xx}$ is vanishingly small, yet, $\kappa_{xx}$ remains finite.

# Heat Conductance of the Quantum Hall Bulk

Ron Aharon Melcer[1,2], Avigail Gil[2], Arup-Kumar Paul[1,2], Priya Tiwary[1,2], Vladimir Umansky[1,2], Moty Heiblum[1,2], Yuval Oreg[2], Ady Stern[2], and Erez Berg[2]

[1] Braun Center for SubMicron Research, Department of Condensed Matter Physics,

Weizmann Institute of Science, Rehovot, Israel, 76100

[2] Department of Condensed Matter Physics, Weizmann Institute of Science, Rehovot, Israel, 76100

## Microscopic model for the heat transport

To calculate the heat current between two thermalized puddles in the bulk, we need to find the energy transfer rate of excitations from one chiral edge state to the other. This requires calculating the transmission amplitude of excitations due to inter-edge interactions, which was derived in [1]. For completeness, we provide the derivation here.

The bosonized Hamiltonian that describes the system of two quantum hall edge states (R and L, see Fig. 4b) interacting through density-density interactions is given by:

$$H = \int dx \frac{v}{4\pi} [(\partial_x \phi_R)^2 + (\partial_x \phi_L)^2] + \frac{u(x)}{2\pi} \partial_x \phi_R \partial_x \phi_L. \tag{S1}$$

For simplicity, we treat the case of integer quantum Hall states, although a similar treatment applies for many fractional states as well. In the integer case, the fields $\phi_R$, $\phi_L$ satisfy the commutation relations:

$$[\phi_R(x), \partial_x \phi_R(x')] = 2\pi i \delta(x - x'),$$
$$[\phi_L(x), \partial_x \phi_L(x')] = -2\pi i \delta(x - x'). \tag{S2}$$

In our model, the inter-edge interaction potential $u(x)$ is taken to be a constant over a finite region of length $L_{int}$ where the puddles are closest to each other, and zero away from this region:

$$u(x) = \begin{cases} u, & 0 < x < L_{int} \\ 0, & \text{otherwise.} \end{cases} \tag{S3}$$

The equation of motion is given by:

$$\partial_t \phi_n(x,t) = i[H, \phi_n] = \pm[v\partial_x \phi_n + u(x)\partial_x \phi_{\bar{n}}], \tag{S4}$$

where $n = R$ or $L$, $\bar{n}$ is the chirality opposite to $n$, and $\pm$ refers to $R$ or $L$. Far away from the interaction region, the velocity of the edge modes is given by $v$, whereas in the interacting region, the renormalized velocity is given by $\tilde{v} = \sqrt{v^2 - u^2}$. By imposing matching boundary conditions for plane wave excitations incoming from one of the edge modes, we find the reflection amplitude, defined as the amplitude of an excitation exiting the interacting region in the same edge state:

$$r(\omega) = \frac{\left(1 + \frac{\tilde{v} - v}{\tilde{v} + v}\right) e^{i\left(\frac{\omega}{\tilde{v}} - \frac{\omega}{v}\right)L_{int}}}{1 + \frac{\tilde{v} - v}{\tilde{v} + v} e^{2i\frac{\omega}{\tilde{v}}L_{int}}}. \tag{S5}$$

Squaring the reflection amplitude to obtain the transmission coefficient of excitations from one edge state to the other:

$$\mathcal{T}(\omega) = 1 - |r(\omega)|^2 = \frac{\frac{u^2}{\tilde{v}^2}\sin^2\left(\frac{\omega}{\tilde{v}}L_{int}\right)}{1 + \frac{u^2}{\tilde{v}^2}\sin^2\left(\frac{\omega}{\tilde{v}}L_{int}\right)} \tag{S6}$$

Assuming that the incoming excitations are thermalized, the thermal current is given by Eq. 3.

**Other possible sources of bulk thermal conductance**

As mentioned in the main text, our experiments cannot directly identify the nature of the low-energy neutral excitations in the bulk that carry the thermal current. Within our proposed model, these neutral excitations originate from the edges of puddles with different filling factors, which are nucleated by disorder. Other sources of neutral excitations are possible: for instance, away from the center of the plateau, the added quasiparticles can form a Wigner crystal, whose low-energy acoustic phonon excitations carry heat. However, it is not obvious how this mechanism can explain the specific temperature dependence of the thermal conductance observed in our experiment. Focusing on the linear response regime, where the temperature difference $\Delta T$ between the Source and PM contacts is smaller than the average temperature, $T$, we found experimentally that the thermal conductance, $\kappa_{xx}T$ is linear in $T$. The bulk thermal conductivity can be written according to Einstein's Relation as $cD$, where $c$ is the specific heat and $D$ is the diffusion constant. The acoustic phonons of a Wigner crystal in a magnetic field and in the presence of long-range Coulomb interactions disperse as [2] $\omega(q) \propto q^{3/2}$. This gives $c(T) \propto T^{4/3}$; and consequently for the temperature dependence of our thermal conductance, the diffusion constant would have to scale as $T^{-1/3}$. It is not obvious which mechanism could lead to such behavior.

Similar reasoning applies to other types of bulk collective modes. For example, Nematic fractional quantum Hall states (observed at certain filling factors [3]) have a linearly dispersing Goldstone collective mode [4]; this mode contributes a term to the specific heat that scales as $T^2$, and consequently, in order to explain our results, the corresponding diffusion constant would have to scale as $T^{-1}$.

**Methodology**

In this section, we describe in detail the method we employed in order to measure the longitudinal thermal conductance in the quantum Hall Effect (QHE) regime. In order to exemplify our technique, we accompany this section with a step-by-step analysis of the raw data.

**Noise measurement**

In the experiment we heat up the source (S) by sourcing opposite sign currents $I, -I$ from contacts $S_1$ and $S_2$ respectively. This causes the dissipation of power $P_S = \frac{I^2}{G_{xy}}$, while keeping the source potential at zero. The dissipated power in the source increases its temperature $T_S$. All the edge modes leaving the source flow to the ground, but in the presence of non-zero longitudinal thermal conductance, heat can propagate through also the bulk and cause an increase to the power-meter's temperature, $T_{PM}$. We simultaneously measure the temperature of the source (S) and the power-meter (PM). This is accomplished by the measurement of the excess current noise (Johnson-Nyquist noise) in amplifiers' contacts located downstream S and PM (contacts $A_S$ and $A_{PM}$ respectively in Fig. 1 of the main text).

The raw noise data for $\nu = 2.05$ ($B = 6.1T$) is plotted in Fig. S1a. The excess current fluctuations are proportional to the temperature of the contact from which the edge emanates (due to the conservation of current). The low-frequency power spectral density of the noise arriving to $A_{\mathrm{S}}$ is given by[5],

$$S_I = 2k_b G^{\star}(T_S - T_0). \qquad (S7)$$

Here, $k_b$ is the Boltzmann constant, $T_0$ is the base temperature and $G^{\star}$ is the effective conductance for noise measurement,

$$G^{\star} = \frac{G_{\mathrm{S \to A}} G_{\mathrm{S \to G}}}{G_{\mathrm{S \to A}} + G_{\mathrm{S \to G}}}, \qquad (S8)$$

with $G_{\mathrm{S \to A}}$ ($G_{\mathrm{S \to G}}$) is the conductance from the source to the amplifier (ground). In the measurement of all the states ($\nu = 2$, $\nu = 7/3$ and $\nu = 5/2$) the region connecting the source to $A_{\mathrm{S}}$ (as well as the region connecting the PM to $A_{\mathrm{PM}}$) was tuned to $\nu = 2$ using a top gate. Thus, we had $G^{\star} = \frac{4}{3}\frac{e^2}{h}, \frac{26}{19}\frac{e^2}{h}$, and $\frac{18}{13}\frac{e^2}{h}$, for $\nu = 2$, $\nu = 7/3$, and $\nu = 5/2$, respectively. Eq. S7 (and the equivalent formula for the PM) is used to extract $T_{\mathrm{S}}$ and $T_{\mathrm{PM}}$ from the measured noise (Fig. S1b).

**Power-meter calibration**

The second step of the measurement is to convert the temperature of the PM to impinging power. This is accomplished by a separate calibration measurement. Here, the source contact is not heated. Instead, we dissipate power directly in the PM contact, by sourcing two opposite currents, $I_{\mathrm{cal}}$, and $-I_{\mathrm{cal}}$ from $S_1^{\mathrm{cal}}$ and $S_2^{\mathrm{cal}}$ (in Fig. 1 of the main text), respectively (Fig. S1c). This allows measuring $T_{\mathrm{PM}}$ against a known power,

$$P_{\mathrm{cal}} = \frac{I_{\mathrm{cal}}^2}{G_{\mathrm{xy}}}, \qquad (S9)$$

(Fig. S1d). Finally, we use the power calibration to convert $T_{\mathrm{PM}}$ to impinging power, which could be plotted as a function of the heater temperature (Fig. S1e), and allows the extraction of $\kappa_{\mathrm{xx}}$.

**S to PM distance dependence**

In order to test the effect of the device geometry on the longitudinal thermal conductance, we performed measurement on devices with different Source-to-Power-Meter (S-PM) distance. First, we reproduced the results of the main text on a separate device, fabricated by a similar design as in Fig. 1, on a different MBE grown GaAs sample. In this device the S-PM distance is $10\mu m$. We then compare the measured longitudinal thermal conductance to a longer device (fabricated on the same material) with $20\mu m$ S-PM distance.

For the fractional states, we observe that the heat flow through the bulk decreases with the S-PM distance (Fig. S2). For $\nu = 5/2$, the longitudinal thermal conductance decreases from $\kappa_{\mathrm{xx}} = 0.37 \pm 0.03\kappa_0$ at 10μm to $\kappa_{\mathrm{xx}} = 0.17 \pm 0.02\kappa_0$ at 20μm. For $\nu = 7/3$, we observe similar behavior as $\kappa_{\mathrm{xx}} = 0.24 \pm 0.01\kappa_0$ for 10μm, and $\kappa_{\mathrm{xx}} = 0.15 \pm 0.01\kappa_0$ for 20μm. While more data is needed in order to determine definitely, our data may suggest that $\kappa_{\mathrm{xx}}$ decays linearly with distance. This ohmic-like behavior is consistent with our theoretical model, since in the model we assume that the temperature is defined "locally" on every puddle. Thus, the transport between the multiple puddles should behave as a "resistors-web" which corresponds to Ohmic conductance in the thermodynamic limit.

This 'resistor like' behavior explains the diminishing impact of the bulk thermal conductance on the previously published works on two-terminal thermal conductance. The typical scale of the two-terminal devices of Refs. [6, 7] was over a hundred microns, and thus we can estimate that the bulk contribution in these much longer devices was below $0.05\kappa_0$ - thus below the resolution of those experiments.

**Two-terminal thermal conductance.**
The measurement of the two-terminal thermal conductance had become a strong tool in the study of QHE states [5, 6, 7, 8, 9]. The measured value of $\kappa_{2T}$ was attributed to the physics of the edge modes, with the bulk neglected. This assumption was justified *posteriori* by the observation of the expected quantized heat conductance. Now, with the direct observation of finite heat conductance through the bulk, we should ask how it affect the two-terminal thermal conductance?

The measured devices are not ideal for such experiments. Namely, the propagation length of heat carried by the edge modes is long while that of the heat carried by the bulk is very short (S-PM distance). Yet, we are able to shed some light on this manner from our current measurements.

The electronic contribution to $\kappa_{2T}$ could be extracted from the relation between the power directly dissipated in the source itself and its temperature [5],

$$P_S = \frac{\kappa_{2T}}{2}\left(T_S^2 - T_0^2\right). \tag{S10}$$

The source-power is plotted against the temperature squared in Fig. S3a for the different values of magnetic field on the $\nu = 2$ plateau. Linear fitting of the low temperature data (up to $T_S = 27\text{mK}$, where the phonon contribution is still negligible[8]) allows to extract $\kappa_{2T}$ according to Eq. S10. $\kappa_{2T}$ is plotted as a function of the magnetic field in Fig. S3b (together with $\kappa_{xx}$ for comparison). In our very short device (S-PM distance of $10\mu m$), it appears that the increase of $\kappa_{xx}$ also causes the two-terminal value to increase (as the contribution of the bulk is added 'in-parallel' to the ballistic edges). However, the actual bulk contribution in long devices, as discussed in the previous section, should become negligible (as in Refs. [6, 7]).

Aside from the non-ideal geometry of our present device for two-terminal measurement, the doping method used here (SPSL[10]), drains energy from the source, which leads to an increase of $\kappa_{2T}$ (this was already observed in our first publication on this subject [7], and analyzed theoretically[11]). As a result, In the current measurement we do-not observe the expected quantized value of $\kappa_{2T}$ at the center of $\nu = 2$ plateau, being $\kappa_{2T} = 3 \times 2\kappa_0$ for the three arms of our device. The excess heat evacuation from the source was $\sim 2\kappa_0$, as expected.

**Additional QHE states**
In addition to the data presented in the main text, we measured the longitudinal thermal conductance of other integer and fractional QHE states. For all the measured states, we were able to detect finite conductance of heat through the bulk. Here, in order to efficiently measure $\kappa_{xx}$ for many values of magnetic field, we did not scan the source temperature. Instead, we relied on $\kappa_{xx}$ being temperature independent and measured for a constant $T_S = 50\text{mK}$. Extracting the longitudinal thermal conductance according to,

$$\kappa_{xx} = 2\frac{P}{T_S^2 - T_0^2}. \tag{S11}$$

We ensure the applicability of this efficient method by the fact that the value of $\kappa_{xx}$ agrees between the two methods (when measured for the same magnetic field).

**Integer QHE $\nu = 3$**

The behavior of this integer state resembles $\nu = 2$ (Fig. S4a). At the center of the plateau $\kappa_{xx}$ vanishes, and it increases as the magnetic field is tuned away from the center. Here, unlike in $\nu = 2$, the value of $\kappa_{xx}$ appears to increase symmetrically in the 'particle' and the 'hole' sides of the plateau.

**Fractional QHE $\nu = 4/3$**

The $\nu = 4/3$ is a robust fractional state in the first Landau level (LL). The spin-up LL is completely full accompanied by the spin-down $\nu = 1/3$ fractional state. Here, similar to the integer fillings, we observe vanishing $\kappa_{xx}$ near the center of the plateau, however $\kappa_{xx}$ increases as the field is tuned away from this value (Fig. S4b). Interestingly, for this fractional state, the region where $\kappa_{xx} \sim 0$ is shifted slightly from the center of the $G_{xx} \sim 0$ region (towards the particle side of the plateau).

**Fractional QHE $\nu = 8/3$**

The $\nu = 8/3$ tends to be more fragile with $G_{xx} > 0$. The order is believed to be identical to the Particle-Hole conjugated $\nu = 2/3$ (on top of a full LL). For this state, we didn't make a thorough study of $\kappa_{xx}$ along the plateau. We did measure $\kappa_{xx}$ in the plateau's center, finding again a finite value $\kappa_{xx} = 0.41 \pm 0.03$ (see Fig. S5c), similar to that of the $\nu = 5/2$ state.

**Further study of $\nu = 2$**

We carefully studied the $\nu = 2$ state, concentrating on the magnetic fields where $\kappa_{xx}$ becomes finite (Fig. S4c). Very interesting is the difference in slope $\partial \kappa_{xx}/\partial B$, which is significantly larger on the particle side (compared to the hole side). This could be attributed to the microscopic realization of the disorder, as the puddles that form when $\nu > 2$ are different from those in the range $\nu < 2$ .

**Raw data**

All the power measurements, which were fitted to extract $\kappa_{xx}$ are presented in Fig. S5.

**Noise acquisition and calibration**

The noise arriving at the amplifier contacts ($A_S$ and $A_{PM}$) is separately amplified by homemade cryo-amplifiers located at the 4.2K plate of the dilution refrigerator. The amplifier's input is connected (in parallel) to the sample resistance, the line capacitance and a superconducting coil, thus forming an *RLC* circuit. In our system the central frequency of the resonant circuit was 632kHz (695kHz) and the bandwidth was 42kHz (44kHz) for $A_S$ ($A_{PM}$). The amplified noise is further amplified by a room temperature amplifier, and is measured with an SRS-865A lock-in amplifier.

The gain of the cryo-amplifiers was calibrated using equilibrium Johnson-Nyquist (JN) noise<Johnson, 1928 #4187;Nyquist, 1928 #4188>. The total measured voltage noise (the spectral density in V²/Hz) could be written as,

$$S_V = A^2 (S_{\text{base}} + S_{\text{JN}}),$$ (S12)

with $S_{\text{base}}$ being the base electronic noise of the amplifier circuit, $A$ the voltage gain, and $S_{\text{JN}}$ the JN noise,

$$S_{\text{JN}} = 4k_b T R ,$$ (S13)

with $R = G_{\text{xy}}^{-1}$. For high enough temperatures (above 30mK), the electronic temperature is well coupled to the cryostat's temperature $T_{\text{C}} = T_0$ ($T_0$ the base electronic temperature) Thus, measuring the equilibrium noise against $T_{\text{C}}$ allows the calibration of the amplification gain, $A$ and the base noise, $S_{\text{base}}$ (Fig. S6). Note that when the cryostat is cooled to the lowest temperatures (below 20mK), the electronic temperature deviates from that of the cryostat, allowing us to determine $T_0$,

$$T_0 = \frac{(S_V - S_{\text{base}})/A^2}{4k_b R}.$$ (S14)

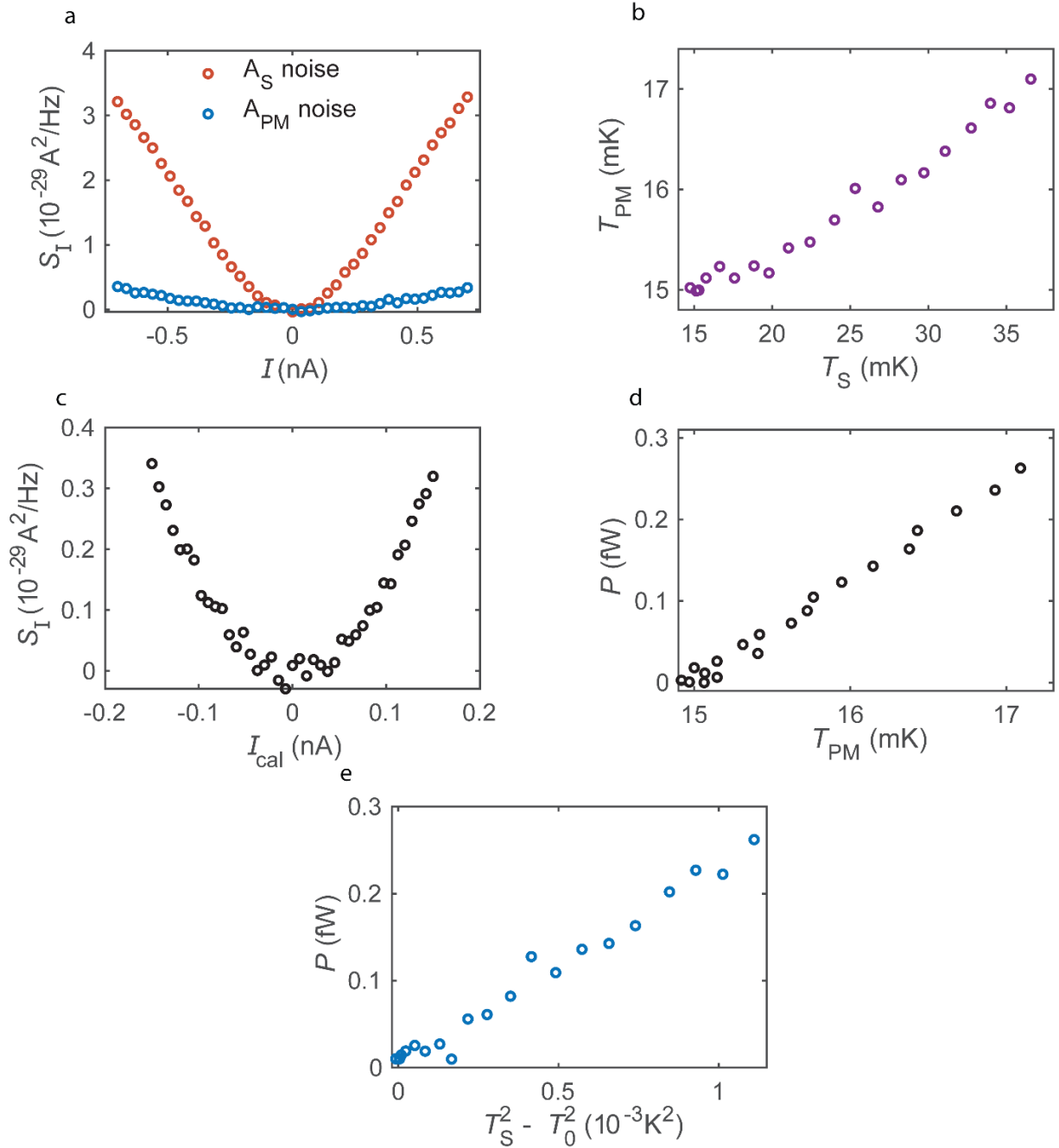

**Fig. S1 – Methodology of power measurement.** Measurement and analysis steps required to measure the heat flow and extract $\kappa_{xx}$. As an example, we present $\nu = 2$ data, measured at $B = 6.1$T, and base temperature of $T_0 = 15$mK. **(a)** Raw noise data. The excess noise measured at $A_S$ and $A_{PM}$ as a function of the sourced current $I$ sourced from $S_1$ (while current $-I$ is simultaneously sourced from $S_2$). **(b)** Power-meter's temperature as a function of the source's temperature extracted from **(a)** using Eq. S7. The heating of the source from $T_0 = 15$mk to a temperature $T_S \sim 40$mK causes the slight increase of the PM's temperature $T_{PM} \sim 17$mK, due to the finite $\kappa_{xx}$. **(c)** Power-meter calibration; raw data. Noise measured at $A_{PM}$ as a function of the direct heating of the PM, by current $I_{cal}$ sourced from $S_1^{cal}$ (while current $-I_{\text{cal}}$ is simultaneously sourced from $S_2^{\text{cal}}$). **(d)** Dissipated power (derived from Eq. S9) as a function of $T_{\text{PM}}$. **(e)** By combining the main measurement **(b)** with the calibration **(d)**, we can plot the power arriving to the PM as a function of the source temperature, and produce the plot presented in the main text Fig. 2a. A linear fit to the power vs. $T_S^2$ gives $\kappa_{\text{xx}}$.

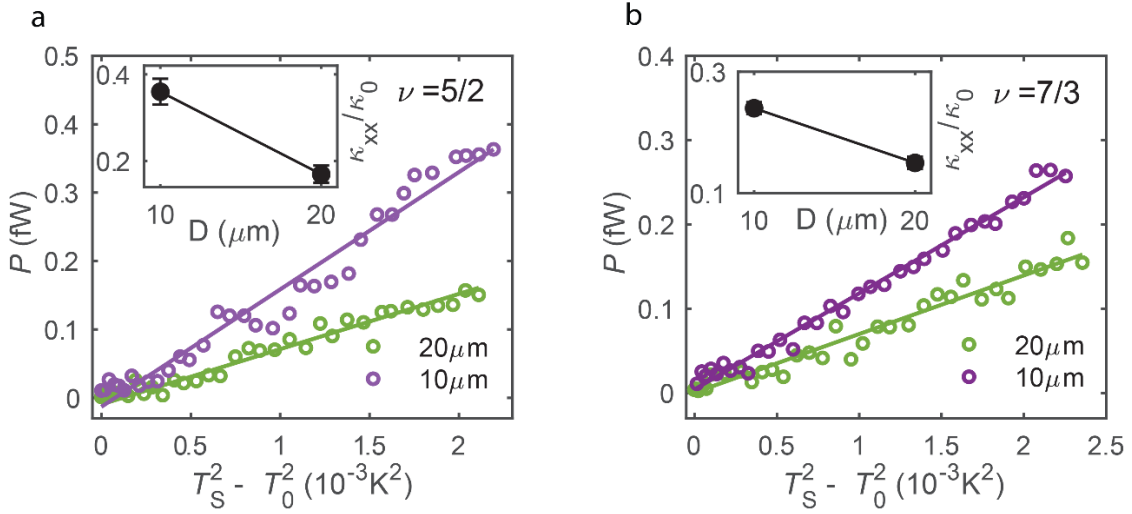

**Fig. S2 $-\kappa_{\text{xx}}$ for different device distance.** The dissipated power in the PM as a function of the source's temperature squared, for two different PM-to-source-distances, D (measured in different devices), $10\mu m$ and $20\ \mu m$. For both fractional states $\nu = 5/2$ **(a)** and $\nu = 7/3$ **(b)**. We observe a decrease of the transferred heat with S-PM distance. This corresponds to $\kappa_{\text{xx}}$ reducing from $\kappa_{\text{xx}} = 0.37 \pm 0.03\kappa_0$ ($\kappa_{\text{xx}} = 0.24 \pm 0.01\kappa_0$) at $10\mu m$ to $\kappa_{\text{xx}} = 0.17 \pm 0.02\kappa_0$ ($\kappa_{\text{xx}} = 0.15 \pm 0.01\kappa_0$) at $20\mu m$ for $\nu = \frac{5}{2}$ ($\nu = 7/3$). Data measured at the plateau center at $T_0 = 10\text{mK}$.

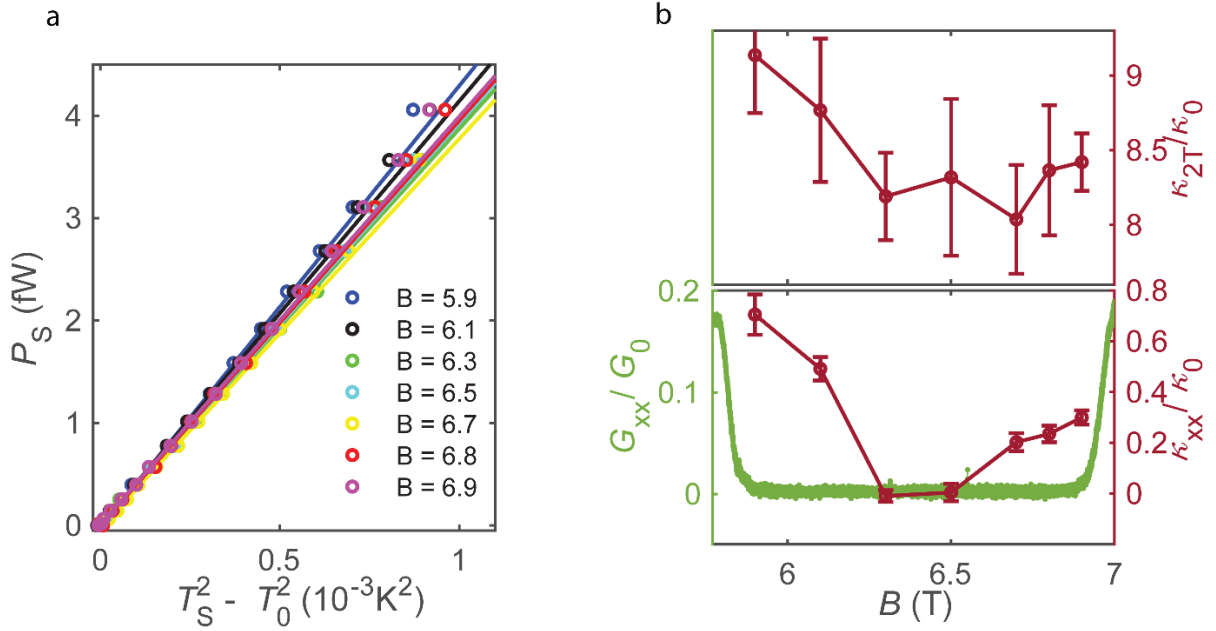

**Fig. S3 – 'Two terminal' thermal conductance on the $\nu = 2$ plateau, at $T_0 = 15\text{mK}$. (a)** Power dissipated at the source, $P_S$, as a function of the source's temperature squared, for different magnetic fields on the $\nu = 2$ plateau. The low temperature data (up to 27mK) is linearly fitted to extract the two-terminal thermal conductance, $\kappa_{2T}$, which changes mildly with magnetic field. **(b) Top-panel** -$\kappa_{2T}$, extracted from **(a)** as a function of the magnetic field (includes $2\kappa_0$ due to donors). **Bottom-panel -** $\kappa_{xx}$, and $G_{xx}$ as a function of the magnetic field (identical to Fig. 2b). It appears that the appearance of finite heat conductance through the bulk causes $\kappa_{2T}$ to increase (in the measured $10\mu m$ S-PM device).

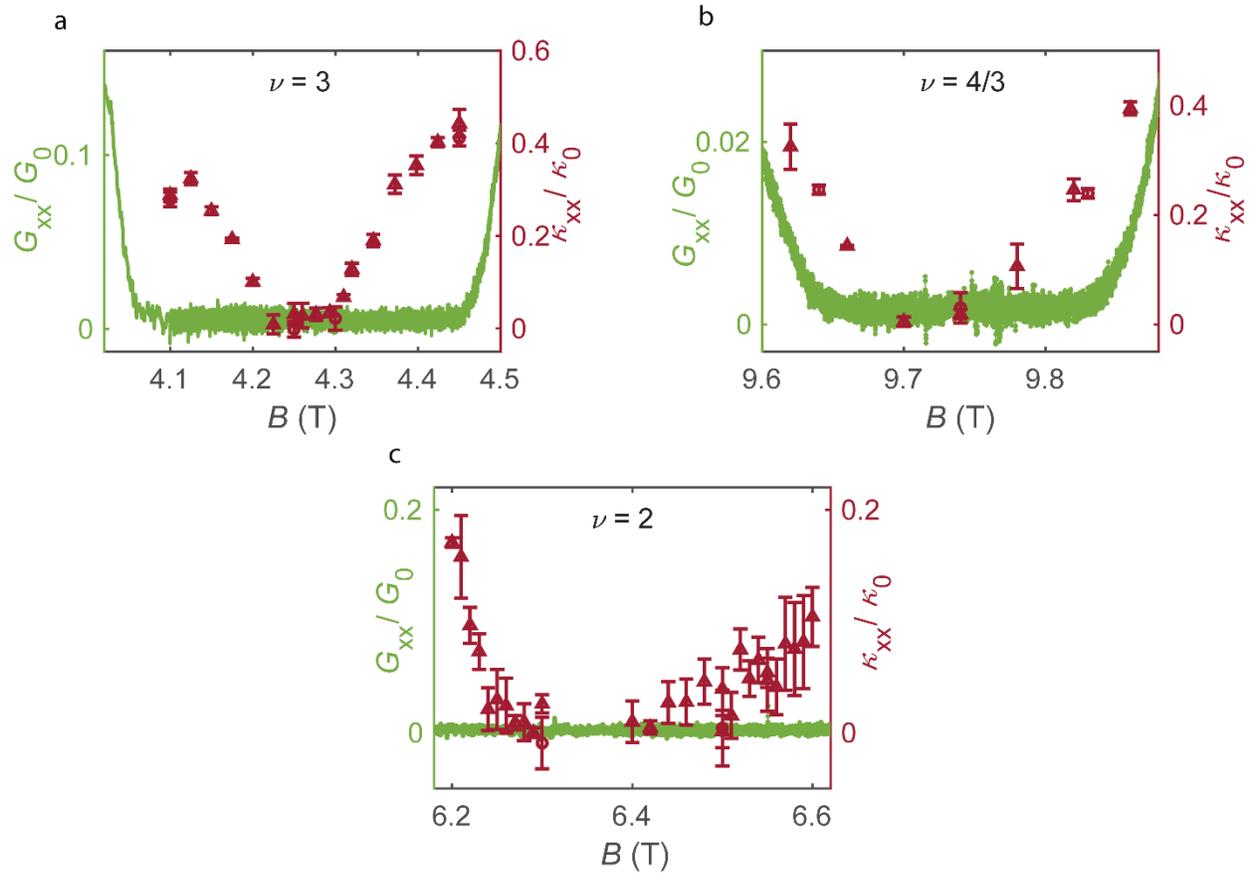

**Fig. S4 – Thermal conductance through the bulk of other QHE states.** Longitudinal electrical conductance (green with scale to the left) and longitudinal thermal conductance (red markers with scale on the right) plotted as a function of magnetic field on the plateaus of **(a)** $\nu = 3$, **(b)** $\nu = 4/3$ and **(c)** $\nu = 2$. The circular markers corresponds to the fitting results of $P$ vs. $T_S^2$ (raw data presented in Fig. S5), and the triangular markers correspond to $\kappa_{xx}$ measured for a single source temperature of $T_S = 50\text{mK}$, and extracted according to Eq. S11.

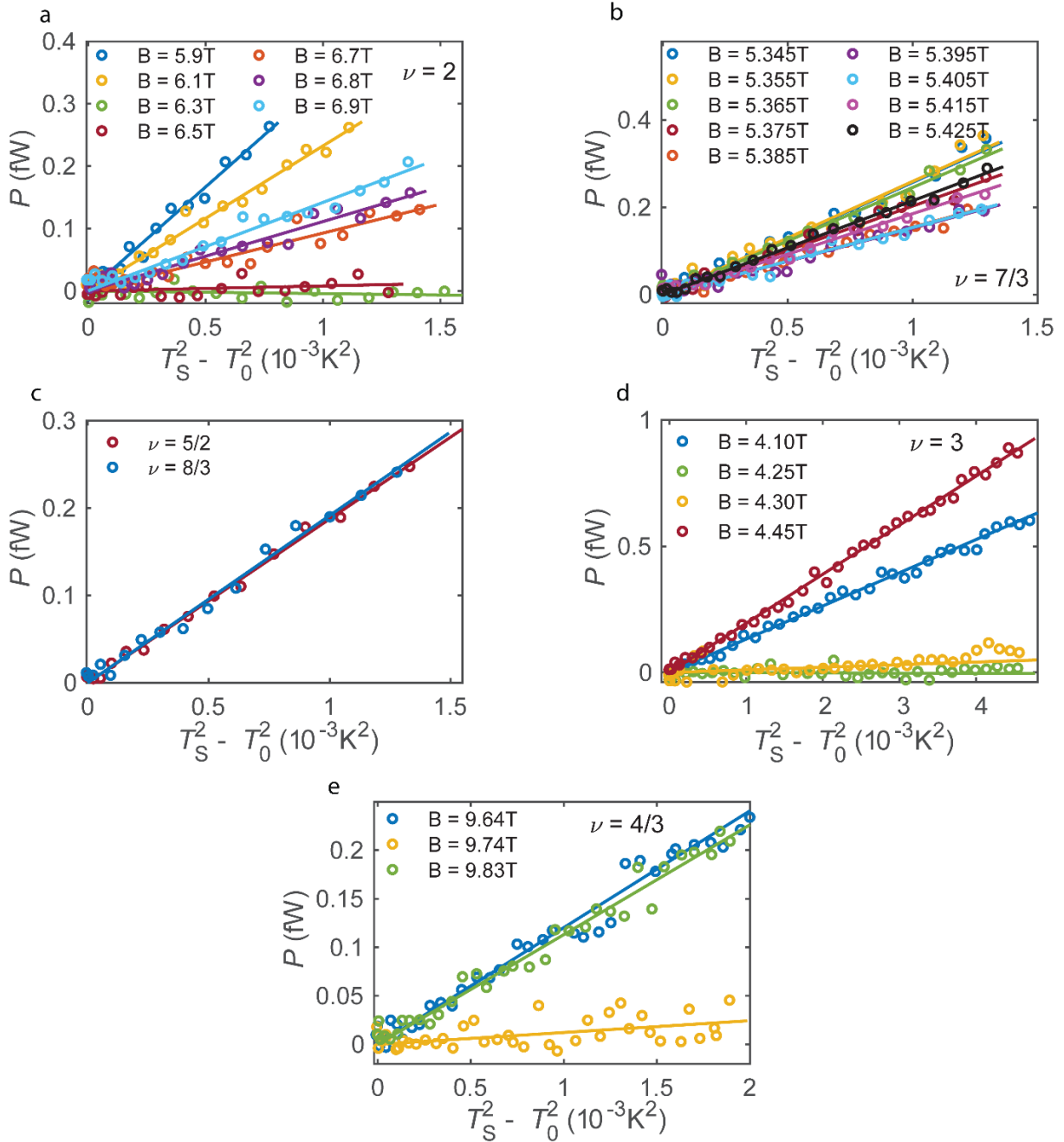

**Fig.S5 – Raw data used to extract $\kappa_{xx}$.** The colored markers represent the power arriving to the PM as a function of the source temperature squared. The data is linearly fitted (colored straight lines) in order to extract $\kappa_{xx}$ (according to Eq. 1 of the main text). Showing the measured data for the results appearing in the main text and the supplementary information: **(a)** $\nu = 2$, **(b)** $\nu = 7/3$, **(c)**, $\nu = 5/2$ and $\nu = 8/3$, **(d)** $\nu = 3$ and **(e)** $\nu = 4/3$.

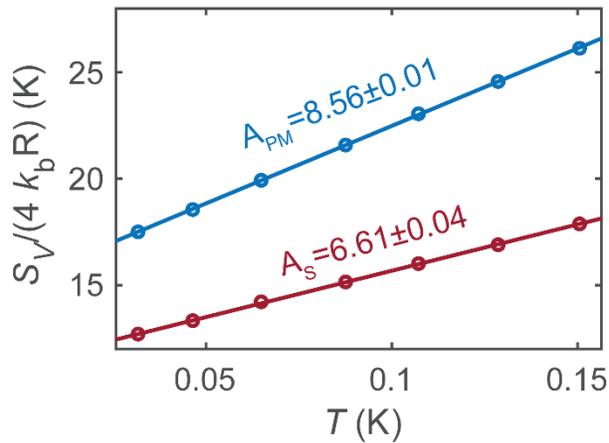

**Fig. S6 – Amplifier calibration.** Equilibrium noise as a function of the cryostat temperature (markers). The noise is linear in temperature, in agreement with the Johnson-Nyquist formula (Eq. S13). This allows us to calibrate the gain according to Eq. S14 (straight lines).